# Body, Clothes, Water and Toys
## -- Media Towards Natural Music Expressions with Digital Sounds --

**Kenji Mase** and **Tomoko Yonezawa**[1]

ATR MIC Research Laboratories

Kyoto, JAPAN

{mase, yone}@mic.atr.co.jp

**ABSTRACT**

In this paper, we introduce our research challenges for creating new musical instruments using everyday-life media with intimate interfaces, such as the self-body, clothes, water and stuffed toys. Various sensor technologies including image processing and general touch sensitive devices are employed to exploit these interaction media. The focus of our effort is to provide user-friendly and enjoyable experiences for new music and sound performances. Multi-modality of musical instruments is explored in each attempt. The degree of controllability in the performance and the richness of expressions are also discussed for each installation.

**INTRODUCTION**

Music is a collection of sounds controlled by musical instruments, which generate constrained sounds in a sophisticated manner. The traditional musical instruments exploit the natural phenomenon of sound generation. The invention of these instruments originated from natural interaction with physical materials. In today's era of artificially generated digital sounds, we propose an approach to provide quasi-natural interaction with digital sounds for the creation of new musical instruments of digital sounds.

We have explored everyday-life materials as interaction media to which we assume to be able to easily relate digital sounds. The relations are called mappings between the interaction and the generated sounds. Three installations, which explore various mappings, are introduced in this paper, i.e., Iamascope+, Tangible Sound 2, and a context-aware music doll.

*Iamascope+(plus)* is an extended version of Iamascope [1], an interactive kaleidoscope that uses a video camera sensor to capture the performer's gesture as the mapping input for music control. It also uses kaleidoscopic graphical images as visual feedback. The Iamascope installation has been successfully exhibited at many places in the world such as SIGRAPH97 in L.A., Ars Electronica Center in Linz, Millennium Expo in London, and Kumano Experience Expo in Japan. We discuss the experiences from the exhibits and extend the original version to give a grater complexity in music control.

*Tangible Sound 2* [2] is an installation that uses water as an interaction medium to control the intuitively appealing feeling of musical flow. Like music, fluids cannot be physically grasped because their shapes are constantly changing. We therefore believe that water has a potential as a suitable interface for performing flowing music. With our instrument, performers interact with water flowing from a faucet into a drain. We have also developed a method for measuring the volume of the water flow and for generating music from this measurement.

The *context-aware music doll* [3] uses musical sounds as it's one and only expressive actuator. The doll is fully equipped with a computer and various sensors such as a camera, microphone, accelerometer, and touch-sensitive sensors to recognize its own situation and the activities of the user, by probing around through multi-modal interaction. The doll has its own internal "mind" states reflecting different situated contexts. The user's multi-modal interaction with the passive doll is translated into musical expressions that depend on the state of mind of the doll.

A case study has told us that the quality of installation in terms of expressive music control heavily depends on the quality of prepared sound mappings designed. The balance between the complexity of sounds and the easiness of interaction has been, and still is, the key. This is what we like to discuss in the workshop based on our experience in prototyping and exhibits. These trials should lead us to the introduction of new modalities in musical instruments, we believe.

In the following sections, we introduce the system configurations of those installations.

**IAMASCOPE+**

Iamascope is an interactive multimedia instrument. Using images captured by a video camera, it generates kaleidoscopic images, which are projected on the wall in front of the performer. The camera shoots the performer, and the generated images consist of the performer, his or her clothes, and anything in the camera view. As a result, Iamascope generates sounds and graphical images as the main performer moves and dances in front of the camera and wall. Iamascope consists of two subsystems, i.e., a vision-to-graphics module and a vision-to-music module, where each runs on an SGI O2 workstation. The vision-to-music module uses a part of the captured images and detects movements of image objects in the view to generate the sounds. The view is divided into small sections to which harmonic musical notes are assigned dynamically. The assignment may be done manually during a performance by an assignment performer or automatically in a pre-defined sequence. Figure 1 shows the view and the system structure of Iamascope.

---

[1] Yonezawa is also a graduate student of Keio University, Graduate School of Media and Governance.

The initial version of Iamascope uses a simplified fixed-function mapping [4] method with which the music sub-module can always generate enjoyable sounds. However, in the automatic mode, the sequence of harmonic codes is pre-defined so that it limits the user's degree of freedom during a performance. Iamascope+(plus) is an extended version to provide the performer with controls for the melody, code and key to some extent.

In the former version, the choice of note is constrained by the current code. Since the code progression timing is predefined in the Iamascope program, the performer has a limited melody expression. In the extension, in contrast, the performer can control the code progression timing by expressive gestures. In addition, the sequence of codes is cyclic in the previous version. The modification provides some control of choice for the code sequence during a performance by the previous verse performance. Lastly, the key can be changed in the modification. The trigger action is a particular gesture that generates a diversion/conversion image display on the screen.

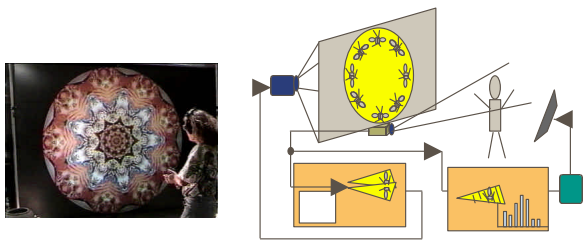

Figure 1: View and structure of Iamascope

## TANGIBLE SOUND 2

Tangible Sound is a musical instrument with a novel user interface based on physical interaction with water. An important aspect of the instrument is that it provides natural tactile feedback when the user touches, scatters and stops the water flow. The spreading water of the instrument is particularly enjoyable, since it is linked to musical tension.

Interacting with water can be a multi-sensorial experience, since it is possible to see, hear and touch the liquid. When we designed our instrument, we realized that tactile interaction was particularly interesting. Hence, if the system is to be used by beginners, it is helpful to create alternative representations for music. Based on our observations, we concluded that tangible water was a very suitable and intuitive musical interface.

### System Design of Tangible Sound 2

Tangible Sound employs the metaphor of "Source and Drains." It is made from the following components: two water tanks, a faucet (acting as the source), and three funnels (acting as drains). The physical layout of these components and its view are shown in Figure 2.

The two tanks are used to generate the flow of water. The first one is on a stand, while the second one is under the stand. The upper tank has a faucet, which makes it possible to regulate the water flow. Within the lower tank, four funnels of different heights serve as drains for the water flow. In addition, a pump is used to transfer water from the lower tank back into the upper tank. The user can divert (touch, scatter) the water, so that some may fall into the main drain (in the lower tank) and some may scatter to the three other drains (funnels). The volume of music is controlled with the faucet attached to the upper tank.

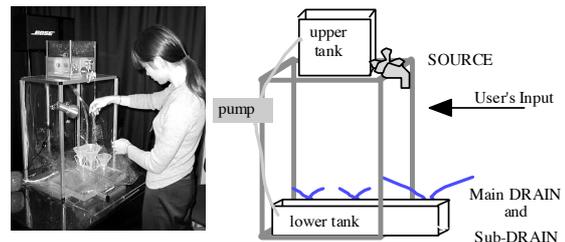

Figure 2: View and structure of Tangible Sound

In order to monitor the water flow in the upper level, we measured its *diameter* under the faucet. Two nichrome wires, measuring the electronic resistance of the water, were used for this purpose. The same technique was used in the lower level funnel. When the volume of the flow exceeds the draining capacity of a funnel, the level of water within that funnel rises. Analog signals are then sent to an i-Cube, a special A/D converter, where the signals are converted into MIDI data, and finally sent to a Macintosh computer via a MIDI interface.

### Music Control Mappings

In combination with the physical configuration of Tangible Sound, we investigated two different software configurations: one for controlling notes and the other for controlling sounds. These software applications were implemented in the MAX/MSP programming language.

*Note Control*

The note control configuration uses the direct input of musical elements with MIDI signal control. We used the QuickTime Instrument as our MIDI sound generator.

This system produces harmonic notes controlled by the flow of water. The *velocity* and *duration* of all MIDI signals are controlled by the upper flow, which the user controls with the faucet. The level of the lower flow generates both *note-on* and *note-number* MIDI signals. For a given funnel, the current level determines the range in which a note is randomly chosen. In other words, if the current level is $x$, then the note number will be chosen between 0 and $2x$.

On the other hand, the *note-on* signals are output either when the water levels in the funnels change or when the drains receive water flow or drops. Discrete notes are generated by the i-Cube every 20 milliseconds. As discussed in more detail later, this gives the impression that every chunk of water corresponds to one note. In addition, every funnel is assigned to one of four harmonies on an eight-tone scale: "E-flat-major," "F-minor," "G-minor" or "A-flat-major." The tone scale corresponds to the height of the drain (e.g., the "E-flat-major" scale is assigned to the highest drain). The user can therefore control four kinds of harmonies by choosing funnels of different heights, where each one expresses a unique level of musical tension.

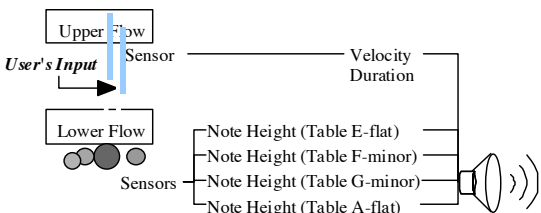

Figure 3: Software structure for note control

*Sonic Control*

The sonic control configuration monitors the drains in order to select *acoustic elements*. In this particular configuration, the user interaction is captured as the *difference* between the upper flow (under the faucet) and the lower flow (in the main drain). The user can touch, stop and scatter the water flow in order to transform "water" sounds into "musical" sounds. Furthermore, the user can change the harmony by pouring water into different funnels.

An audio sample of flowing water (recorded in a bathroom) is used as a basis to generate the audio output. The MIDI output is composed of five tracks, in which the flowing water sample is played in loops. Time lapses are introduced so that the sample is not played concurrently in different tracks. When the user touches the water flow, the sound is continuously filtered to resonant frequencies of a musical harmony. In other words, "water" sounds are transformed into "musical sounds" according to the following procedure:

Every track is associated with one frequency. In other words, the output of the five tracks is a harmony. In every track, the sample sound continuously changes into a band-limited, musical sound by resonance. The frequency of this sound is the same as the track frequency. Every drain is associated with one of four harmonies, namely "E-flat-major," "F-minor," "G-minor" or "A-flat-major". The height of the drain corresponds to the height of the harmony. When water is poured in a drain, the five sound tracks continuously change to match the harmony associated with this drain.

**CONTEXT-AWARE MUSIC DOLL**

We developed a music doll to provide a non-verbal communications system that can sense various non-verbal actions of the user and the doll and translate them into music. The unique and novel part of the system is that the translation is performed based on the doll's embedded internal character. The doll displays its own built-in autonomous behaviours when responding to various external approaches. At the same time, it becomes a musical communications media for its partner, providing the user with some musical expression controls.

**Sensor Doll**

The sensor-doll contains a card-sized, full specification Windows PC with Wave-LAN (wireless network adapter) ethernet networking capability and various sensors. The sensor values and recognized gesture data are sent to the sound generating module in the MIDI format. The MIDI signals are interpreted and processed to control the internal state automaton and sound and music synthesis. The sounds and music are sent out to the room's loud speakers as well as the doll's internal wireless loudspeaker. The room's loudspeakers play ambient music while the wireless speaker plays the doll's voice. For the wireless link, we use an FM transmitter and an off-the-shelf FM radio.

The sensor-doll contains the following components; a small PC (Intel Pentium II 333MHz HD, 128MB memory, Win2K OS, dimensions = 140mm x 100mm x 40mm), wireless network card, power (battery, 7.2V, 2360mAh), A/D signal converter, and many kinds of sensors attached to the shell of the plush bear-like doll. Concerning the sensors, there is a USB camera in the doll's nose, a USB microphone in its right ear, a G-force sensor (gyro sensor) at the center of its body, four bend sensors in each arm and leg, five touch sensors (in both hands, the head, belly, and back), infra-red proximity sensors (at the bottom of its hip and nose), and heat sensors near the PC and outside the body. The infrared proximity sensors are used to judge the doll's distance from objects such as tables, doll's chair, and people. The heat sensors are used to measure the temperature difference between the outside and inside. The dimensions of the doll are 30cm tall (standing), 28cm tall (sitting), 46cm around the waist, and 1500g (including the battery). The system runs for approximately 30 minutes with the 2360mAh battery in the full functional mode. It also has an extra connector to get AC-DC converted power for longer use in experiments. Figure 4 shows a snapshot of its use.

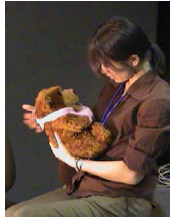

Figure 4: A Snapshot of Playing with the Music-Doll

Based on the sensor installations described above, the doll can detect and recognize various events to control its state transitions. For instance, the proximity sensor on the hip can detect whether the doll is sitting or not, the touch and bend sensors in each hand can detect the hand-shaking gesture, and the combination of the touch sensor in the back and proximity sensor in the face can detect the state of being held by a human partner. The G-force sensor can be used as additional information to support the detection of events by giving reliability indices.

The sensor inputs are first sent to the pattern recognition module, where events and gestures are detected for use in the decision of the doll's internal state and forwarded to the context-dependent gesture interpreter (CGI) module. The CGI module plays the role of outputting the internal state automaton; there are five internal states that are denoted by Interaction Levels. It collects information on (i) the internal state, (ii) recognized events and gestures, and (iii) raw sensor data, in order to generate corresponding behaviors and responses, which are then mapped to sound and musical elements in the music mapping module. Finally, it generates and outputs sounds through the synthesizer.

**Context-dependent Musical Expressions**

Sound and music outputs are the only actuators for the sensor-doll to express its internal moods and responses against its partner's inputs. We carefully designed and prepared the musical elements, and then mapped them depending on the interpreted contexts. Some expressions are real-time responses to inputs and others are autonomous displays of the doll's state. The current number of prepared mappings is approximately 30, most of which are used at Interaction Level 3. At level 3, the doll partially performs as a musical controller allowing its partner to play music with it. The sound and musical elements and the controls are listed below. Examples of music mappings are described in Table 1. Consequently, the same input to the doll can result in different expressions appearing, depending on the context.

Global controls - global loudness, harmony, key, and tempo.
Breath sound controls - interval, loudness, resonance filter intensity, and structure of the harmony.
Voice sound controls - loudness, filtering frequency, speed, and delay time
Melody controls – musical notes, length, and loudness.
Rhythm controls - loudness and pattern.

**Table 1:** Examples of Music Mappings and Controls

| Gesture (Sensor) | Interaction Level | Music or Sound Mappings and Controls |
|---|---|---|
| Hold Hands | 2 | Harmony Structure |
|  | 3 | Melody Loudness |
| Bend Left Hand | 1 | Voice Frequency |
|  | 2 | Note Table |
|  | 3 | Melody Note |

We also employ a rhythm detector to detect rhythmical inputs during Interaction Level 2. It uses the inputs as triggers to move to level 3, exploiting the captured rhythms as basic tempo units for the accompanying musical performance at level 3, i.e., the musical communications state.

**CONCLUSION**

The works described in this paper aim to highlighting the importance of the musical multi-modality in music performance; i.e., the visual representations, touch sensitivity to water and plush toy, full body action such as hugging and dancing, personality of instrument, etc. The computational analysis and exploitation of these modality are the next step.

**ACKNOWLEDGMENTS**

We thank Sidney Fels, Kazushi Nishimoto, Brian Clarkson, Tim Chen, and Axel Mulder for their contributions and collaboration for the work performed in Department 2, ATR MIC, which provide the important fundamentals of this paper. We also thank Ryohei Nakatsu and Michiaki Yasumura for their support in this work.